# Enhancing Supply Chain Transparency in Emerging Economies Using Online Contents and LLMs

1st Bohan Jin
Peking University
Guanghua School of Mangement
Beijing China
bohanjin@stu.pku.edu.cn

2nd Qianyou Sun
Peking University
Guanghua School of Mangement
Beijing China
qysun@stu.pku.edu.cn

3rd Lihua Chen*
Peking University
Guanghua School of Mangement
Beijing China
chenlh@gsm.pku.edu.cn

*Abstract*—In the current global economy, supply chain transparency plays a pivotal role in ensuring this security by enabling companies to monitor supplier performance and fostering accountability and responsibility. Despite the advancements in supply chain relationship datasets like Bloomberg and FactSet, supply chain transparency remains a significant challenge in emerging economies due to issues such as information asymmetry and institutional gaps in regulation.

This study proposes a novel approach to enhance supply chain transparency in emerging economies by leveraging online content and large language models (LLMs). We develop a Supply Chain Knowledge Graph Mining System that integrates advanced LLMs with web crawler technology to automatically collect and analyze supply chain information. The system's effectiveness is validated through a case study focusing on the semiconductor supply chain, a domain that has recently gained significant attention due to supply chain risks.

Our results demonstrate that the proposed system provides greater applicability for emerging economies, such as mainland China, complementing the data gaps in existing datasets. However, challenges including the accurate estimation of monetary and material flows, the handling of time series data, synonyms disambiguation, and mitigating biases from online contents still remains. Future research should focus on addressing these issues to further enhance the system's capabilities and broaden its application to other emerging economies and industries.

## I. INTRODUCTION

In the context of globalization facing unprecedented challenges such as the COVID-19 pandemic, trade wars, and the Russia-Ukraine conflict, the security of supply chains has become increasingly critical.

Ensuring a resilient and secure supply chain is essential not only for individual businesses but also for national economies, and supply chain transparency plays a pivotal role in safeguarding this security. Transparency ensures that companies can monitor their suppliers' performance, from sourcing raw materials to delivering products, fostering accountability and responsibility throughout the supply chain[1]. Moreover, supply chain transparency enhances firms' and governments' ability to manage risks and respond to changes by driving knowledge creation and mediating the relationship between flexibility and risk management[2]. news about supply chain information provides a useful instrument for analyzers, facilitating the estimation of company performance and equity value[3].

Nowadays, supply chain relationship datasets like Bloomberg, FactSet and Mergent are already helping to enhance supply chain transparency around the world. Bloomberg Terminal's Supply Chain Analysis (SPLC) dataset is one of the most accessible supply chain relationship datasets. Primarily based on publicly listed companies under the U.S. Securities and Exchange Commission's S-K regulations, this dataset also leverages information from sources such as earnings call transcripts, capital markets presentations, and press releases. Beside presenting supply chain relationships, the SPLC dataset also provides details like cost types, transaction amounts, and the percentage of total sales represented by these transactions. A non-public proprietary algorithm is employed to estimate these figures[4]. Other two major supply chain relationship data provider, FactSet and Mergent, also includes strategic partners associated with key companies through investments, joint ventures, research collaborations, and integrated product offerings. However, the number of companies they cover are less than that of Bloomberg SPLC's[5]. These datasets jointly provide a useful tool for enterprises and governments to promote supply chain transparency, thus controls the supply chains risks.

Despite the advancements in datasets, supply chain transparency remains a significant challenge in many emerging economies. For instance, in countries like Brazil, institutional gaps in regulation severely undermine the ability to maintain transparency across supply chains. These issues make it difficult to enforce standards and trace goods and services effectively, posing risks to both local and global stakeholders[6]. In China, another typical emerging economy, although it is stated in *Standards for the Contents and Formats of Information Disclosure by Companies Publicly Offering Securities No. 2—Contents and Formats of Annual Reports* that public companies must disclose the proportion of annual sales attributable to their top five customers and the proportion of annual procurement attributable to their top five suppliers, and that companies are encouraged to disclose the specific names of these suppliers and customers, their procurement or sales amounts, only about 20% of companies disclosing the names of their suppliers and customers, accounting for a very small section of the whole economy[7]. This makes the companies in emerging economies much less covered in mainstream supply chain relationship datasets like Bloomberg and FactSet.

Information asymmetry is a prevalent issue in the supply chains of emerging economies damaging supply chain transparency[8]. Without a workable disclosure mechanism, both businesses and governments often only have partial knowledge of the supply chain, such as their direct suppliers, customers, or entities within their jurisdiction. This lack of comprehensive, real-time, and accurate insight into the broader supply chain landscape limits the effectiveness of risk management and decision-making. Existing statistical systems also fall short of providing robust supply chain data, leaving gaps in understanding supplier relationships. Therefore, it is still hard to ensure supply chain transparency in emerging economies[9].

Knowledge graphs (KG) are a type of non-SQL database widely used for knowledge base construction and search engine optimization[10]. By applying graph analysis algorithms, knowledge graphs can identify critical entities, supporting automated risk identification and proactive risk management[11]. Furthermore, they enhance supply chain reasoning by fully utilizing the knowledge embedded in nodes and relationships. This approach improves the effectiveness of supply chain analysis compared to traditional methods, as is evidenced in the automotive industry[12].

In recent years, large language models (LLMs) have significantly enhanced the ability to extract effective knowledge from diverse and heterogeneous information sources, owing to their remarkable generalization capabilities[13]. This advancement facilitates the integration of disparate data sources, thereby creating conditions for the construction of supply relationship datasets for emerging economies and improving the transparency of their supply chains.

This study proposes a construction scheme for a Supply Chain Knowledge Graph Mining System based on large language models and knowledge graphs, aimed at automating the collection of supply chain information from online contents. To validate the effectiveness of this scheme, our research focuses on the semiconductor supply chain, a domain that has recently garnered significant attention due to supply chain risks[14]. A prototype of the system and its interactive interface was successfully developed. Comparisons with traditional datasets reveal that this system demonstrates greater applicability for emerging economies such as mainland China, providing a viable pathway to enhance supply chain transparency.

## II. RELATED WORKS

Recently, natural language processing technology has seen rapid development, enabling researchers to automatically extract valuable information from vast amounts of text. As far as we know, the first trial to collect business information from online contents is done by[15], who used Jaccard coefficient to infer the relationship of business entities on the internet based on the concept of co-occurrence. Their way of fabricating the right keyword for search engine is still inspiring for our research. [16] used semi-supervised statistical inference method to tackle Named Entity Recognition (NER) of business entities in online contents. Later, A team from Toshiba Corporation used DeepDive, a machine learning system based on Markov Logic Network (MLN) to extract company relationship from web content, attaining a precision of 67% for cooperative relations and 81% for competitive relations[17]. The first study dealing with automatically generating supply chain information is done by Pascal et al.[18], utilizing a deep learning model, BiLSTM, to extract supplier-buyer relationships from a public corpus through a traditional NER (Named Entity Recognition) and RE (Relation Extraction) pipeline. Since OpenAI team proposed ChatGPT[19], the use of large language models (LLMs) in information extraction (IE) has become prevalent[20]. This IE method has become increasingly continent after mainstream LLMs propose JSON mode for structured output[21]. Some study report good performance[22], while comparisons between small language models (SLMs) and LLMs indicate that LLMs often lag behind SLMs but excel at tackling complex problems[23] and show much potential in IE of universal context[24]. A recent study done by [25] attempted to map supply chain relationships using information extracted by LLMs from Wikipedia, achieving 0.82 accuracy in the relationship extraction task, thus demonstrating the applicability of LLMs in enhancing supply chain transparency. However, all studies on AI-driven supply chain relationship information extraction have not compared their results with existing datasets like Bloomberg, nor have they discussed the applicability of their method in the context of emerging markets. Our study has built a Supply Chain Knowledge Graph Mining System to address these gaps.

## III. SYSTEM DESGIN

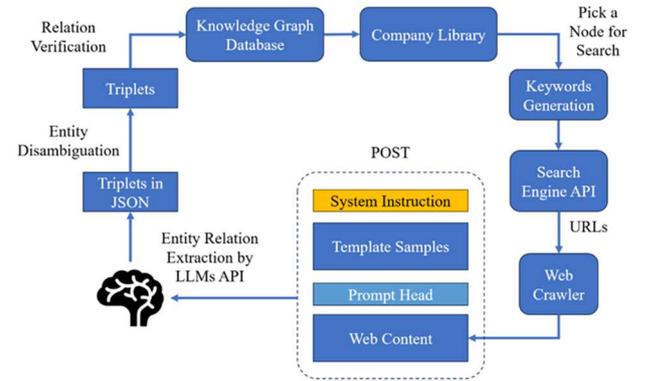

Fig. 1. Supply Chain Knowledge Graph Construction Algorithm Based on LLMs

As shown in Fig. 1, we introduced a Supply Chain Knowledge Graph Mining System based on LLMs. Our system addresses a critical challenge of mining company entities and entity relationships in a wide range of Internet texts and accurately constructs a supply chain knowledge graph. The system will first use industry research reports as the seeds to set up the Company Library, take the companies that appear in the research reports as the entry point, automatically search for information from the Internet, and use LLM to extract supply chain company entities and their relationships to build a supply chain network. It will repeat the cycle until no new company entities are mined. The specific process of the algorithm is introduced below.

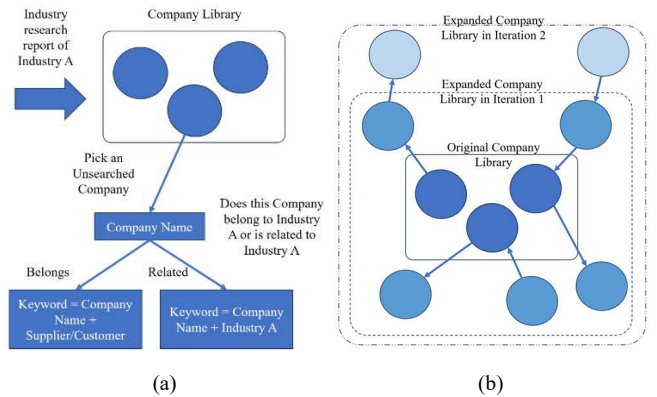

(a)      (b)

Fig. 2. (a) The process of keywords generation for search engine API. (b) the way our system walkthrough the supply chain network

## A. Collection of Public Information

To obtain multilingual textual data related to corporate supply chains, this study employs a combination of localized documents and internet text. The localized documents primarily consist of industry research reports, which consists of most of the names of important companies in the industry we focus on. This process ensures the Company Library is initialized and ready to generate the keywords we need in the next step. The detailed way how keywords are generated is described in Fig. 2(a). Then, internet text is gathered through an iterative discovery process. By utilizing search engine APIs, relevant pages are retrieved and crawled, from which suppliers and customers are extracted. This process is then repeated for the identified suppliers and customers, gradually enabling a deep walkthrough of the entire supply chain, as is shown in Fig. 2(b).

## B. Entity Relation Extraction

This study utilizes large natural language models (LLMs) to achieve joint extraction of knowledge graph triplet, that is, entities and relationships. Through a few-shot approach and iterative optimization of prompt engineering, the model is guided to return entities, relationships and their attributes in JSON format, which are subsequently loaded into a knowledge graph database. The detailed content of prompt is demonstrated in Fig. 3. Considering the Template Samples are too long and vary with different industries, we only provide the framework in the picture. To ensure traceability of information, the model is required to provide textual evidence for its judgments within the JSON, whose existence within the document is validated by code. Moreover, considering the benefit of efficiency, sharing contextual understanding [26] and reducing error propagation[27], we let the LLMs do end-to-end extraction by giving out entity and relationship at the same time. This technology enables us to automate the generation of supply chain knowledge graphs with minimal human intervention.

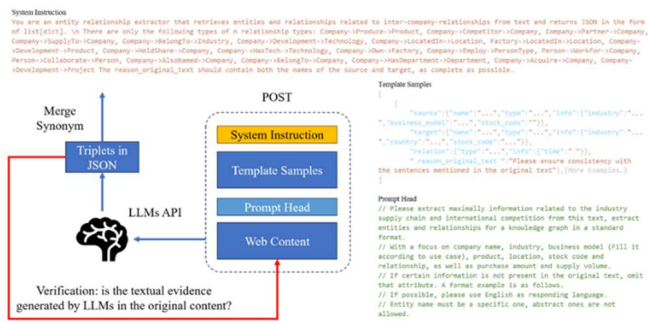

Fig. 3. the way Supply Chain Knowledge Graph Mining System interact with LLMs

## C. Entity Relationship Disambiguation

At the node level, discrepancies in the representation of the same entity across different textual data often lead to the creation of multiple nodes for the same entity in the initial knowledge graph, such as "Huawei", "Huawei Technologies Co., Ltd." and "HUAWEI." To address this, as is shown in Fig. 4, we adopt Match based on Relation and Match based on Embedding Similarity to preliminarily find suspected synonyms, and then use LLMs to testify whether they are really synonyms. Once synonym recognition is completed, the capabilities of the company name management system will be gradually enhanced by maintaining a synonym list on a SQL database, facilitating more accurate and faster synonym detection in the future. Since wrong synonym detection result can heavily affect the structure of network, the synonym list is double checked by human.

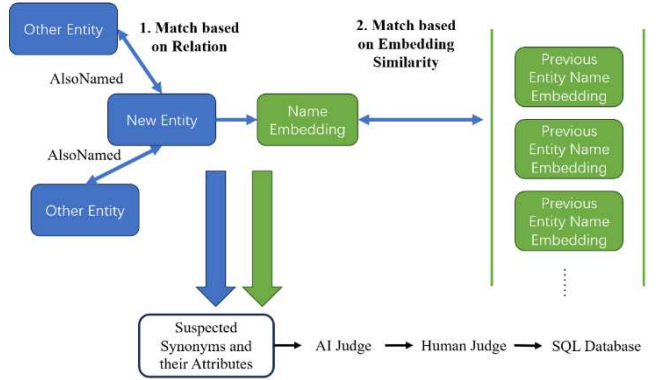

Fig. 4. Synonym discovery process

At the relationship level, as LLMs can easily produce misclassifications and reversed supply chain direction during batch extraction of relationships, we need to construct a discriminative model for secondary verification of each generated relationship one at a time, thereby improving data accuracy. We also use LLMs to attain this by utilizing the features of entities and relationships, raising the precision by 8%.

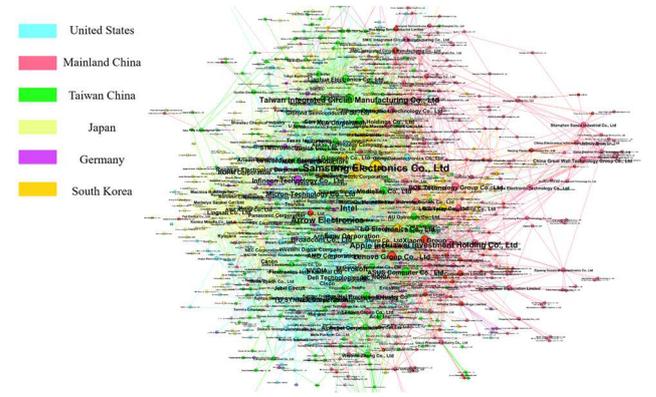

Fig. 5. Semiconductor supply chain constructed with Bloomberg SPLC dataset

## IV. EVALUATION METHOD

To compare the data collected by the Supply Chain Knowledge Graph Mining System developed in this study with existing datasets, we selected the Bloomberg SPLC database, which is widely utilized in international journals and holds a strong reputation within the industry, as our benchmark. From October to December 2023, we manually downloaded information on the 1st tier and 2nd tier suppliers as well as customers of the top 100 companies by market capitalization in the global semiconductor industry from the Bloomberg terminal, obtaining a benchmark dataset comprised of 1,753 companies and 5,749 supply relationships, as is shown in Fig. 5. Due to the limitations of the Bloomberg terminal, we were only able to download information on the top 20 suppliers and customers for each company.

Then, using the previously mentioned Supply Chain Knowledge Graph Mining System, we constructed a semiconductor supply chain knowledge graph comprising over 300,000 nodes and 640,000 relationships. This graph includes approximately 20,000 company nodes and 9,226 supply chain relationships (i.e., Company-Supply-Company relationships). After validation through a discriminative model, we extracted 6,613 reliable supply relationships from the initially screened relationship data, of which 4,484 are unique. Theoretically the system can go on to collect more supply chain information, but considering the size of our dataset should match that of the Bloomberg benchmark dataset's, we only used this dataset composed of around 5,000 valid relationships.

To evaluate the performance of our model, we use the measurement of Precision, because our model is running on an open internet environment, we mainly focus on whether the supply chain relationship collected is true, instead of whether the model can collect all the supply relationships of the company. Precision is determined by Eq. (1).

$$Precision = \frac{True\ Positive}{True\ Positive + False\ Positive} \quad (1)$$

We randomly sampled 200 supply chain relationships collected by the system, and use human-labelled result as the true value to calculate Precision.

To further quantify and compare the characteristics of the Bloomberg SPLC dataset and the dataset acquired from our Supply Chain Knowledge Graph Mining System, we employed three metrics from social network theory: degree, graph density, and modularity. The degree indicates the number of edges a node has within the network. Graph density represents the proportion of existing edges (R) to all possible edges, calculated using Eq. (2).

$$D = \frac{2R}{N(N-1)} \quad (2)$$

where N is the number of nodes in the network. Modularity characterizes the tendency of nodes within a network to connect with each other in submodules, forming relatively isolated parts[28], calculated by Eq. (3).

$$Q = \frac{1}{2R} \sum_{i,j} \left( A_{ij} - \frac{k_i k_j}{2R} \right) \cdot \delta(s_i, s_j) \quad (3)$$

where A is the adjacency matrix, $A_{ij}$ is a binary variable indicating the presence of an edge between nodes i and j, k is the degree of the nodes, and $\delta(s_i, s_j)$ is a function that indicates whether nodes i and j are in the same module. A high modularity in the supply relationship network may suggest the formation of relatively isolated supply chains among companies, with minimal overlap, or it may result from data incompleteness that disrupts the network. To further compare the applicability of the two datasets for enterprises in emerging economies, we also calculated the metrics for a subnet comprising only Chinese mainland companies. The calculation is done in an opensource network analysis software Gephi.

## V. RESULT

In terms of performance, the supply chain relationship collected by our system achieves a 77% precision after validation through a discriminative model and shows better performance than Bloomberg for emerging economies.

Through alignment with the aforementioned Bloomberg SPLC dataset, it was found that among the 1,753 companies in the downloaded Bloomberg SPLC dataset, 1,020 were covered by our Supply Chain Knowledge Graph Mining System, indicating a relatively similar scope of node coverage. However, among the 5,749 relationships within the Bloomberg SPLC dataset, only 173 were duplicated in the relationships identified by the Supply Chain Knowledge Graph Mining System. This suggests a considerable disparity in the scope of supply relationship coverage between the two databases.

TABLE I. COMPARISON OF GRAPH MEASUREMENTS

|  | Economic | Average Degree | Graph/Sub-graph Density | Graph/Sub-graph Modularity |
|---|---|---|---|---|
| **Bloomberg SPLC** | All | 6.559 | 0.0019 | 0.439 |
|  | US | 10.42 | 0.0075 | 0.375 |
|  | PRC | 5.427 | 0.0033 | 0.65 |
| **Supply Chain Knowledge Graph Mining System** | All | 2.485 | 0.0004 | 0.774 |
|  | US | 4.093 | 0.0019 | 0.563 |
|  | PRC | 4.224 | 0.0017 | 0.67 |

From TABLE I. we can observe that the data extracted by our Supply Chain Knowledge Graph Mining System has lower average degree, lower graph density, and higher modularity compared to the Bloomberg SPLC dataset, indicating a sparser network structure, indicating a wider range of data collection. The sparsity itself is not a big problem because running the system for a longer period can easily collect more information and make up for the sparsity. In this paragraph, we focus more on the comparison between All Data, Mainland China (noted as PRC in TABLE I. ) and U.S. In the Bloomberg SPLC dataset, the graph density for Mainland China is notably lower than the U.S., while its modularity is higher than the U.S. and the whole network, leading to an illusion that the semiconductor supply chain of Mainland China is sparse and has already been decoupled from the majority of the world. In contrast, the data extracted by the Supply Chain Knowledge Graph Mining System does not show such a gap in graph density for Mainland China; instead, its modularity is lower than the overall level and the average degree is even higher than that of the U.S. Comparison of Fig. 5 and Fig. 6 also reveals that, in the Bloomberg SPLC dataset, companies from Mainland China are more dispersed at the periphery of the network with fewer internal connections within the economic entity. Conversely, in the data extracted by the Supply Chain Knowledge Graph

Mining System, companies from Mainland China exhibit tighter internal connections, forming a cohesive cluster.

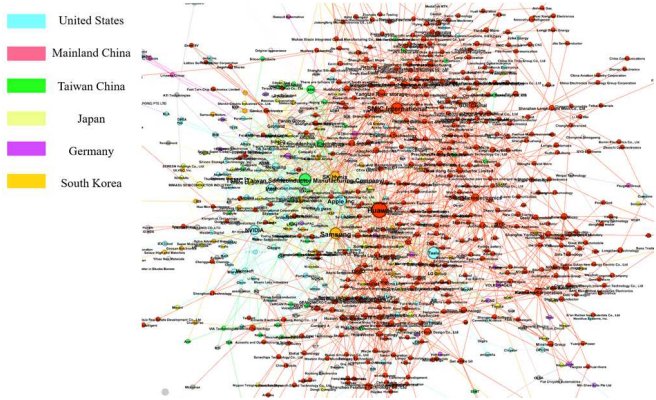

Fig. 6. Semiconductor supply chain constructed with our Supply Chain Knowledge Graph Mining System

Moreover, 143 Malasia companies and 189 India companies are also included in this dataset, among which 54 Malasia semiconductor supply chain relationships and 76 India's are detected. These characteristics highlight better applicability of Supply Chain Knowledge Graph Mining System to enterprises in emerging economies, capable of more comprehensively demonstrating the supply chain structure, thus providing a valuable supplement to existing datasets.

Additionally, the system has also collected other types of relationships, such as those related to competition and cooperation, technology, personnel, equity, and factory establishment, as shown in TABLE II. These leave more potential for future researches.

TABLE II. THE NUMBER OF RELATIONSHIPS OF EACH TYPE COLLECTED BY OUR SYSTEM

| Relationship Type | Count |
|---|---|
| Company-Produce-Product | 41976 |
| Company-Competitor-Company | 10903 |
| Company-Supply-Company | 9226 |
| Company-Partner-Company | 7133 |
| Company-HasTech-Technology | 4335 |
| Company-Develop-Product | 4330 |
| Company-HoldShare-Company | 3896 |
| Person-WorkFor-Company | 3381 |
| Company-LocatedIn-Location | 3129 |
| Company-Own-Factory | 1429 |
| Person-Collaborate-Person | 1328 |
| Product-Use-Technology | 992 |

## VI. DISCUSSION

In this study, we proved the applicability of LLMs in supply chain information extraction problem, and the same method can be utilized to extract many more types of relationship from the open-source contents at very little marginal cost. In our Supply Chain Knowledge Graph Mining System, LLMs do not only shows great potential in extracting triplets of knowledge graph, but also many auxiliary works like entity disambiguation and relationship verification, forming most parts of the pipeline. This enables researchers to do information extraction job much more efficient than before. Moreover, our system gathers online information by choosing the keyword for web search dynamically, gradually exploring the whole supply chain from the core to periphery, instead of processing all the text from a pre- predetermined range of corpus. This method has not been used in previous studies and is especially useful to supply chain information extraction. Last but not least, our study uses network metrics to prove better coverage of our system to companies in economies like mainland China and India than Bloomberg SPLC, demonstrating our system's great potential in enhancing supply chain transparency in emerging economies. Considering the poor situation of supply chain visibility in emerging economies, the result of this study is a significant supplement to the current supply chain database, and leaves opportunity for more future studies in the supply chain structure of emerging economies.

Since supply chain information extraction is a recently emerging field, there are not many previous works for us to compare. The most recent published work reports a 0.72 F1 score by BiLSTM model, a 0.42 Precision rate in "A supplies B" judges, and a 0.33 in "B supplies A" judges[18]. A preprint article [25] also reports a 0.71 accuracy (which we believe should be precision) in supply chain relationship RE task. Our work achieves a precision of 0.77 in supply chain relationship extraction, reaching a satisfying outcome. Moreover, we are able to let the LLMs generate contextual evidence based on the original content, promoting the accuracy of information extraction and enhancing fact traceability. Nevertheless, our study uses network measurements to prove the better coverage of mainland China companies of our system than Bloomberg SPLC, demonstrating the system's great potential in enhancing supply chain transparency in emerging economies. However, we recognize that there is still considerable scope for exploration.

First, the Bloomberg SPLC dataset typically provides estimates of the monetary value involved in supply chain relationships. This information is crucial in supply chain analysis as it helps researchers identify which supply chains are more significant and allows them to assign weights to the edges in network analysis. Achieving this level of detail remains challenging in the current system, as most online content does not disclose such specific financial data, unlike the annual reports of publicly traded companies. Fortunately, graph neural networks (GNNs) and material flow analysis (MFA) provide robust technical tools to infer the volume of money or goods flowing through the links in a supply chain network, guiding the next steps of our research.

Second, the ability to perform time series analysis on a dataset is often crucial for researchers. While the publication date provides a temporal indicator for when a supply chain relationship occurred, the content itself may refer to events that predate the publication time. Moreover, online content is not always periodic; a supply chain relationship might be reported once in 2020 and not mentioned again for several subsequent years. In such cases, it is unclear whether the relationship has continued or been disrupted. More mathematical models are necessary to address these issues.

Lastly, our study focuses on mainland China and the semiconductor industry because China is the largest emerging economy and manufacturing hub, and semiconductors are a prominent topic in the media of two highly digitized countries, China and the United States. This focus provides a wealth of online content for our research. However, whether the same information extraction strategies are effective for other emerging economies and industries, particularly those that receive less media coverage, remains an open question. Additionally, media reports may be biased, leading to some companies, countries, and industries receiving more attention and thus generating more available data than others. This can introduce biases into the data retrieved by the Supply Chain Knowledge Graph Mining System. Future studies should address these challenges.

## VII. Conclusion

Supply chain transparency is essential for enhancing the resilience and security of businesses worldwide, particularly in emerging economies. This study underscores the potential of utilizing online content and large language models (LLMs) to improve the visibility and understanding of supply chain network structures. By integrating advanced LLMs with internet crawling technology, our Supply Chain Knowledge Graph Mining System effectively extracts supply chain relationship information, providing valuable insights for decision-makers. The system's applicability in emerging economies, such as mainland China, complements the data gaps in existing supply chain datasets like Bloomberg SPLC, highlighting its potential to address the challenges of supply chain transparency in these markets. Future work is needed to enhance the information extraction of monetary and material flows, time series robustness, and to mitigate biases in the dataset.